\documentclass[%
 reprint,
 amsmath,amssymb,
 aps,
 superscriptaddress,
]{revtex4-2}

\usepackage{graphicx}
\usepackage{dcolumn}
\usepackage{physics}
\usepackage{booktabs}
\usepackage{amsmath,bm}
\usepackage[colorlinks=true,linkcolor=black,anchorcolor=black,citecolor=black,filecolor=black,menucolor=black,runcolor=black,urlcolor=black]{hyperref}

\begin{document}

\preprint{APS/123-QED}

\title{Study of the magnetoelastic effect in nickel and cobalt thin films at GHz range using X-ray microscopy}

\author{Marc Rovirola}
\email{marc.rovirola@ub.edu}
\affiliation{Dept. of Condensed Matter Physics, University of Barcelona, 08028 Barcelona, Spain}
\affiliation{Institute of Nanoscience and Nanotechnology (IN2UB), University of Barcelona, 08028 Barcelona, Spain}
\author{M. Waqas Khaliq}
\email{mkhaliq@cells.es}
\affiliation{Dept. of Condensed Matter Physics, University of Barcelona, 08028 Barcelona, Spain}
\affiliation{ALBA Synchrotron Light Source, 08290 Cerdanyola del Vallès, Spain}
\author{Travis Gustafson}
\affiliation{{Department of Physics, Norwegian University of Science and Technology (NTNU), 7034 Trondheim, Norway}}
\author{Fiona Sosa}
\affiliation{Dept. of Condensed Matter Physics, University of Barcelona, 08028 Barcelona, Spain}
\author{Blai Casals}
\affiliation{Institute of Nanoscience and Nanotechnology (IN2UB), University of Barcelona, 08028 Barcelona, Spain}
\affiliation{Dept. of Applied Physics, University of Barcelona, 08028 Barcelona, Spain}
\author{Joan Manel Hernàndez}
\affiliation{Dept. of Condensed Matter Physics, University of Barcelona, 08028 Barcelona, Spain}
\affiliation{Institute of Nanoscience and Nanotechnology (IN2UB), University of Barcelona, 08028 Barcelona, Spain}
\author{Sandra Ruiz-Gómez}
\affiliation{Max Planck Institute for Chemical Physics of Solids, 01187-Dresden, Germany}
\author{Miguel Angel Niño}
\affiliation{ALBA Synchrotron Light Source, 08290 Cerdanyola del Vallès, Spain}
\author{Lucía Aballe}
\affiliation{ALBA Synchrotron Light Source, 08290 Cerdanyola del Vallès, Spain}
\author{Alberto Hernández-Mínguez}
\affiliation{Paul-Drude-Institut für Festkörperelektronik, Leibniz-Institut im Forschungsverbund Berlin e.V., 10117 Berlin, Germany}
\author{Michael Foerster}
\affiliation{ALBA Synchrotron Light Source, 08290 Cerdanyola del Vallès, Spain}
\author{Ferran Macià}
\email{ferran.macia@ub.edu}
\affiliation{Dept. of Condensed Matter Physics, University of Barcelona, 08028 Barcelona, Spain}
\affiliation{Institute of Nanoscience and Nanotechnology (IN2UB), University of Barcelona, 08028 Barcelona, Spain}

\date{\today}

\begin{abstract}
We use surface acoustic waves of 1 and 3 GHz in hybrid piezoelectric-magnetic systems with either cobalt or nickel as a magnetic layer to generate magnetoacoustic waves and directly image them using stroboscopic X-ray magnetic circular dichroism imaging. Our measurements visualize and quantify the amplitudes of both acoustic and magnetic components of the magnetoacoustic waves, which are generated in the ferromagnetic layer and can propagate over millimeter distances. Additionally, we quantifiedy the magnetoelastic strain component for cobalt and nickel through micromagnetic simulations. We obtained a drop in the magnetoacoustic signal at 3 GHz suggesting a speed limit for the efficient magnetoelastic coupling in our hybrid devices.
\end{abstract}

\maketitle



\section{\label{sec:intro} Introduction}
There is an increasing interest in manipulating magnetization dynamics to create low-power consumption devices capable of transmitting and encoding information efficiently \cite{locatelliSpintorqueBuildingBlocks2014}. A challenge lies in effectively creating and controlling magnetic excitations in nanodevices. The most common way to manipulate magnetization typically involves the use of magnetic fields. However, this approach has several downsides, including issues with non-locality, and challenges with integrating controllable magnetic fields into micrometer-scale devices. Other approaches explore interactions of conducting electron's spin \cite{RALPH20081190, brataasCurrentinducedTorquesMagnetic2012}, lattice phonons \cite{Bukharaev2018straintornicRev,Foerster_2019}, or ultrashort light pulses \cite{Mangin2014} with magnetic excitations. Promising alternatives have been proposed, such as the use of strain. The field of straintronics \cite{Bukharaev2018straintornicRev} studies strain-induced effects in solids and special attention is directed to magnetic materials as a promising alternative for reduced energy-consumption devices \cite{Barangi2015StraintronicsPowerEff}. Strain couples to magnetic states through the magnetoelastic (ME) effect, which is the change of magnetic properties due to mechanical deformation. This effect has already been proposed and used in many applications \cite{Foerster_2019} including the reversing magnetization of patterned nano-magnets \cite{wang2014magrev_strain,Huang2014magrev_strain} or inducing a phase transition from antiferromagnetic ordering to ferromagnetic ordering \cite{cenkerReversibleStraininducedMagnetic2022}. 

The ME effect can be used to generate spin waves---collective excitations of magnetic order---through surface acoustic waves (SAWs) \cite{gangulyMagnetoelasticSurfaceWaves1976,fengMechanismInteractionSurface1982}, which are strain waves propagating at the surface of a solid. SAWs generate an oscillating strain in both space and time which can create an oscillating magnetic anisotropy field on a magnetostrictive material \cite{yang2021acoustic,puebla2022perspectives}. This dynamic anisotropy field may induce a magnetization variation with the same wavelength and frequency as the SAWs and the excitation may propagate up to millimeter distances \cite{casalsGenerationImagingMagnetoacoustic2020}. These hybrid waves are referred to as magnetoacoustic waves (MAWs) and have been studied in a wide range of materials by measuring acoustic attenuation as a function of the magnetic state \cite{weiler2011elastically,gowthamTravelingSurfaceSpinwave2015b,Labanowski2016,seemann2022magnetoelastic,Kuszewski_2018} or by direct imaging using magneto-optic Kerr effect (MOKE) \cite{Kuszewski_2018,McCord_AEM_2022} or X-ray magnetic circular dichroism (XMCD) \cite{casalsGenerationImagingMagnetoacoustic2020,rovirola2023_physrevapp}. Using the same principles, more recently also Néel Vector waves have been observed in antiferromagnetic CuMnAs \cite{khaliq2023antiferromagnetic} by X-ray linear dichroism (XMLD).

SAWs are typically generated by means of interdigital transducers (IDTs) deposited on piezoelectric substrates. IDTs are interlocking arrays of metallic electrodes that can excite SAWs up to GHz frequencies. IDTs are used in electronics as delay lines and filters and their integration into micrometric devices is well-established \cite{SAW_roadmap_2019}. SAW-based technology can be efficiently integrated into low-power devices since the SAW amplitude depends on the amplitude of the oscillating voltage applied to the IDT instead of the current.

Magnetic imaging techniques based on MOKE \cite{Zhang2020MOKEspinwaves,qinNanoscaleMagnonicFabryPerot2021} or XMCD \cite{foerster2017direct,casalsGenerationImagingMagnetoacoustic2020, rovirola2023_physrevapp} bring the opportunity to directly measure MAWs and thus study the effect of the coupling between strain and spin waves both in space and in time. Recent experiments have shown that the ME effect is as efficient in dynamic processes up to 500 MHz as is in static processes \cite{foerster2017direct}, thus fueling the idea of using SAWs in magnonic applications. One of the key questions in this field is whether the ME remains efficient at higher frequencies, which are characteristic of magnetic resonances.

In this paper, we study MAWs at 1 and 3 GHz in nickel and cobalt thin films using stroboscopic X-ray photoemission electron microscopy (XPEEM) combined with XMCD imaging. Our analysis reveals that both nickel and cobalt exhibit MAWs across a wide range of magnetic fields, each material displaying different profiles and amplitudes. With the help of micromagnetic simulations, we are able to quantify the strength of the magnetoacoustic coupling for both magnetic materials at the studied frequencies.

\section{Experimental methods}\label{section:expmet}
\begin{figure*}[ht]
    \centering
    \includegraphics[width=0.9\textwidth]{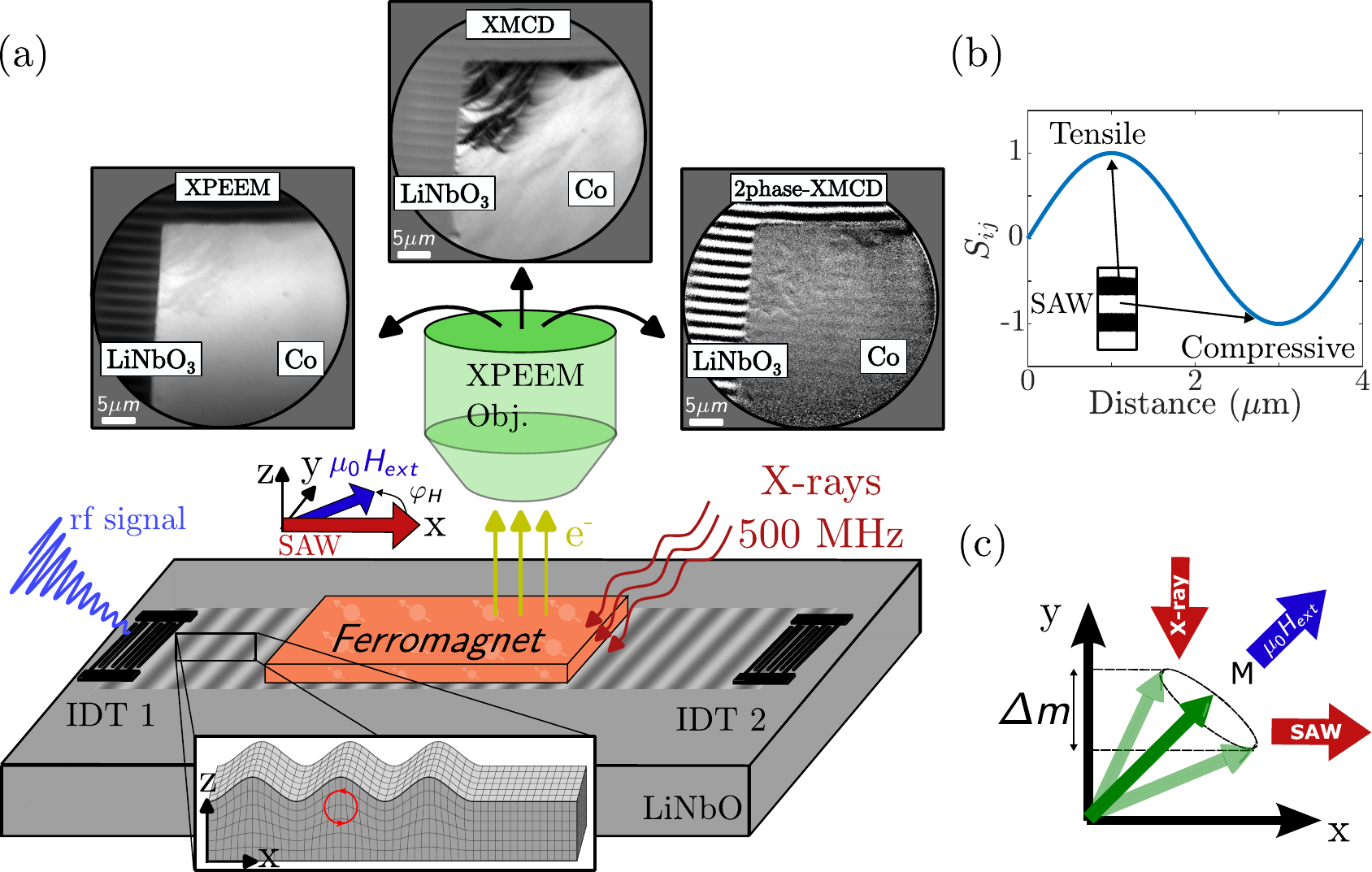}
    \caption{ \textbf{(a)} Schematic of the experimental setup. The thin film of either nickel or cobalt is grown in the middle of the acoustic path. The cobalt sample can only be excited at 1 GHz by either IDT. The microscope (XPEEM Objective) is operated at 10 kV with respect to the sample to accelerate the electrons excited by the X-rays. From this setup, we can obtain XPEEM and XMCD images. The magnetic field is applied in-plane at an angle $\varphi_{H} = 45^\circ$ relative to the SAW propagation direction. \textbf{(b)} Normalized schematics of SAW strain phases, compressive and tensile. The difference in piezoelectric voltage between each phase is labeled as $V_{pp}$, from which the strain components can be determined. \textbf{(c)} Representation of the direction of SAW, magnetic field, and X-ray. In this experiment, the SAW and X-ray are perpendicular to each other. The horizontal dashed lines represent the change in magnetization projection captured with the X-rays.
    }
    \label{setup}
\end{figure*}

\begin{figure}[ht]
    \centering
    \includegraphics[width=0.9\columnwidth]{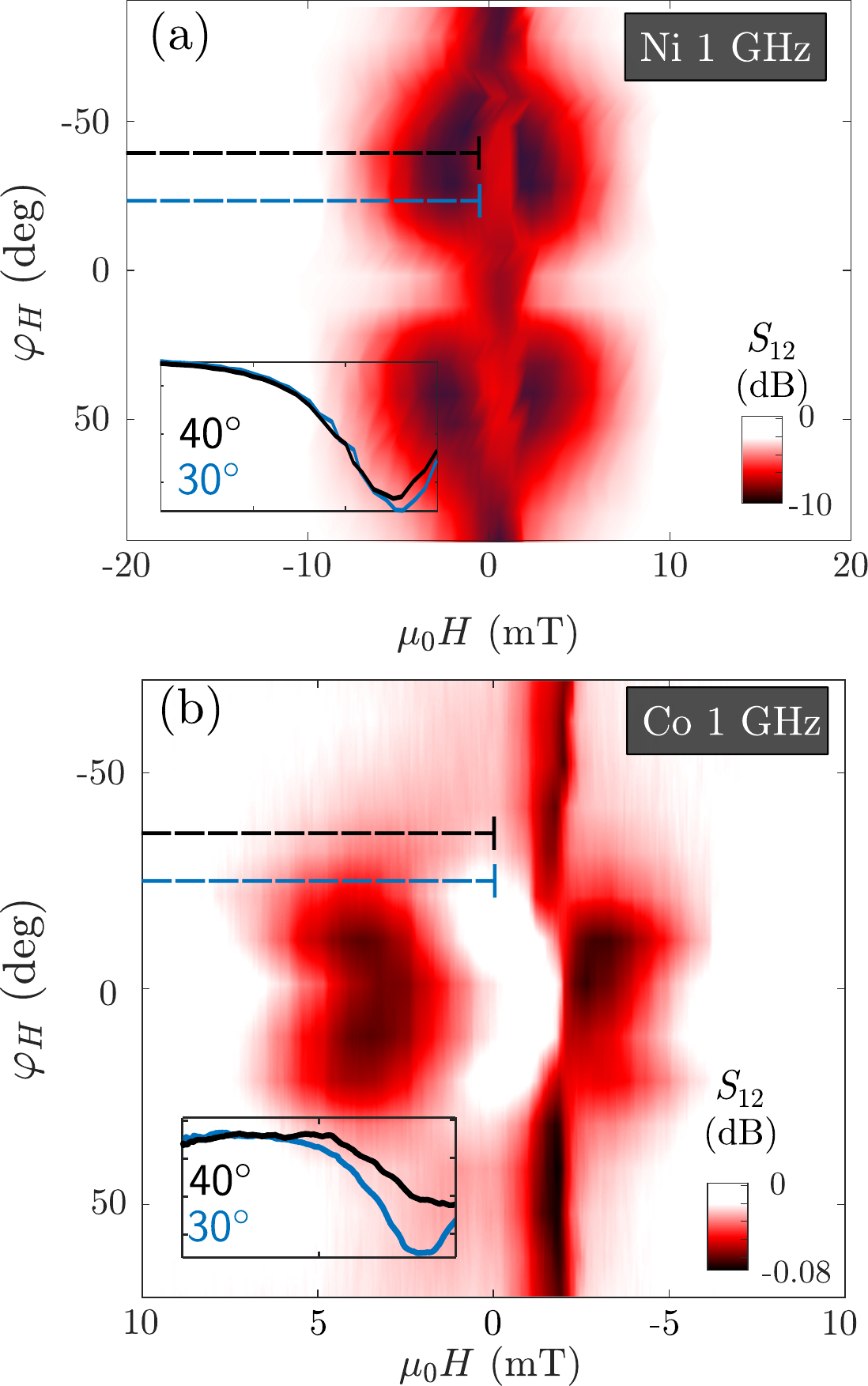}
    \caption{
    \textbf{(a)} Nickel $S_{12}$ attenuation at 1 GHz at several angles between the SAW and the magnetic field showing different absorption values with a maximum close to 45$^\circ$ and minimums at 0$^\circ$ and above 60$^\circ$. The inset shows the S12 attenuation at positive magnetic fields of the angles 40$^\circ$ (black curve) and 30$^\circ$ (blue curve).
    \textbf{(b)} Cobalt $S_{12}$ attenuation at 1 GHz for several angles between the SAW and the magnetic field. Cobalt shows absorption when the SAW and the magnetic field are at an angle between -30$^\circ$ and 30$^\circ$. The inset shows the S$_12$ attenuation at positive magnetic fields of the angles 40$^\circ$ (black curve) and 30$^\circ$ (blue curve).
    }
    \label{NiCoFMR}
\end{figure}
Thin films of nickel (Ni) and cobalt (Co) with a thickness of 20 nm were grown on the acoustic path of several SAW delay lines by e-beam evaporation. The delay lines consist of pairs of unidirectional IDTs deposited on piezoelectric lithium niobate (128Y-cut LiNbO3) substrates, see Fig.\ \ref{setup} \cite{foersterSubnanosecondMagnetizationDynamics2018}. The IDTs are designed to generate SAWs with frequencies of 1 GHz and 3 GHz for the Ni samples and 1 GHz for the Co ones. These frequencies are conveniently selected to match the synchrotron repetition rate (499.654 MHz), thus allowing for stroboscopic images. To generate SAWs, we excite the IDTs with an oscillating electrical signal of the appropriate frequency that is transformed into strain through the inverse piezoelectric effect \cite{gangulyMagnetoelasticSurfaceWaves1976}. Notice that SAWs are limited to a substrate depth on the order of the SAW wavelength \cite{Slobodnik1976_SAW}, corresponding to 4 um and 1.32 um for 1 and 3 GHz, respectively. Within this depth region, the SAW causes a periodic modulation of the crystal lattice in the substrate and on any thin film deposited on its surface. The SAW propagating through the hybrid device generates a periodic effective magnetic field in the ferromagnetic film as a result of the ME effect.

The experiment was performed at the XPEEM experimental station of the ALBA Synchrotron Light Facility \cite{AballeALBA2015}. A schematic representation of the setup is shown in Fig. \ref{setup}(a). Details on synchronized SAW stroboscopic XPEEM measurements and the upgraded setup to support $>500$ MHz excitations can be found elsewhere \cite{foerster2019quantification, Waqas2023}. Since the metallicity of Ni and Co shields the piezoelectric field of the LiNbO3 substrate, only changes in magnetization can be observed in the areas covered by the ferromagnetic films. To resolve these changes, two XPEEM images with different light helicity, circular right and circular left, were subtracted to obtain an XMCD image, in which the contrast variations are directly related to the in-plane magnetization of the samples. The insets of Fig. \ref{setup}(a) display the three imaging modalities of the setup. The top-left corner depicts an XPEEM image with clear SAWs contrast on the LiNbO$_3$, accompanied by a weak contrast of magnetic domains on the Co. The top-center depicts an XMCD image that highlights the magnetic contrast on the Co magnetic domains, accompanied by faint traces of residual SAWs on the LiNbO$_3$ due to thermal drifts in subsequent images (and thus different SAW propagation velocities). The top-right corner depicts a 2-phase-XMCD image, which consists of capturing two XPEEM images with a 180-phase shift in SAW for each X-ray helicity, followed by a subtraction between both phases and helicities. This method enhanced both the magnetic and piezoelectric contrast while eliminating static contrast such as magnetic domains, enabling a clearer measurement of MAWs. Figure \ref{setup}(b) shows the two opposite phases of the SAW, tensile and compressive. Figure \ref{setup}(c) shows the directions of the SAW, the magnetic field, and the X-rays. The SAW and the magnetic field are at 45$^\circ$ with respect to each other to exert maximum magnetoelastic torque, whereas the SAW and X-rays are perpendicular to each other in order to detect the magnetization changes in the $y$-direction as represented by the dashed lines.



\section{Results}\label{section:results}

\begin{figure*}[ht]
    \centering
    \includegraphics[width=1.9\columnwidth]{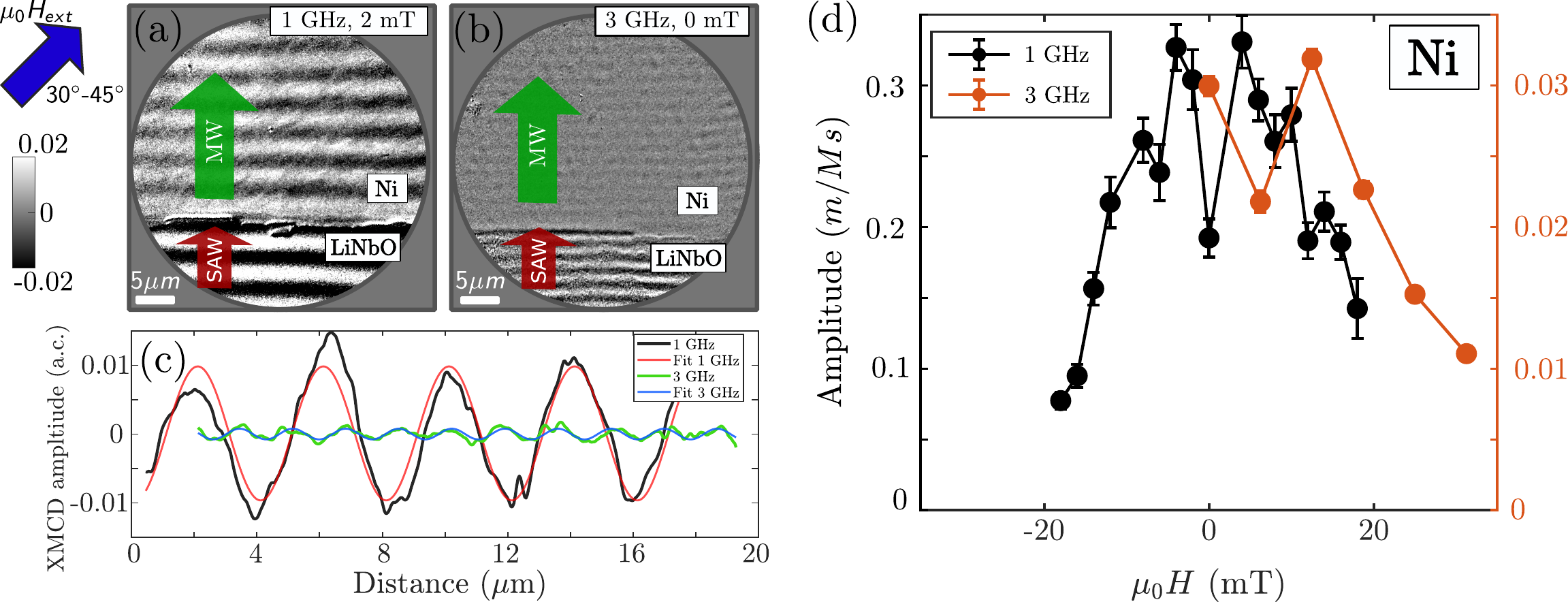}
    \caption{
    \textbf{(a)} XMCD image of Ni showing the MAW driven at 1 GHz.
    \textbf{(b)} XMCD image of Ni showing the MAW driven at 3 GHz.
    \textbf{(c)} Experimentally determined MAW profiles at 1 GHz (black curve) and 3 GHz (green curve). the red and blue curves are fittings to a sinusoidal function.
    \textbf{(d)} Dependence of the MAW amplitude on the external magnetic field. The left axis represents the excitation at 1 GHz, while the right axis represents the excitation at 3 GHz. 
    }
    \label{NiSWAmp}
\end{figure*}

An initial characterization of the samples was performed to study the angular dependence of their acoustic-ferromagnetic resonance (a-FMR) at 1 GHz. The experiment consists in placing our hybrid device (see, Fig.\ \ref{setup}(a)) between the poles of an electromagnet and measuring the transmitted acoustic signal, $S_{12}$, with a vector network analyzer at different angles between the SAWs and the in-plane magnetic field. In contrast to FMR, where an rf-magnetic field is used to drive the magnetization into resonance, a-FMR uses the periodic strain field of the SAW to induce magnetization precession by modulating the effective magnetic field of the sample. Additionally, the non-zero wave vector of the SAW has to be taken into account in the determination of the resonant magnetic fields and frequencies \cite{gowthamTravelingSurfaceSpinwave2015b}. 

The coupling between magnetization and SAWs can generate MAWs under certain conditions of the external magnetic field. At resonance, significant energy is transferred from the SAW to the magnetic system, resulting in SAW attenuation and phase shift that are detected in the S21 measurement \cite{weiler2011elastically}. The SAW attenuation results are presented in Fig.\ \ref{NiCoFMR} as a function of the applied field magnitude and angle (in both cases, the magnetic field was swept from negative to positive values). In both samples, a large attenuation is observed, regardless of the angle, at a fixed magnetic field that corresponds to the magnetization-switching field of the film. In addition, Ni (Fig. \ref{NiCoFMR}(a)) shows the typical four-fold shape with large attenuation between $\pm$30$^\circ$ and $\pm$50$^\circ$, and a maximum around 45$^\circ$ where the ME torque is the largest \cite{Puebla_2020, weiler2011elastically, dreherSurfaceAcousticWave2012}. As the angle between SAW and the magnetic field exceeds 50$^\circ$ or drops below 30$^\circ$, the attenuation decreases, reaching the minimum ME torque at 0$^\circ$ and 90$^\circ$. In contrast, the SAW attenuation for Co (Fig. \ref{NiCoFMR}(b)) is strongest at magnetic field angles -30$^\circ$ and 30$^\circ$, above which the attenuation rapidly decays. Similar results for Co can be found in Ref. \cite{spinPumpingCoWeiler2012}.

Prior to capturing XMCD images of the ferromagnetic sample, the IDTs are tested by imaging the piezoelectric component of SAWs on the LiNbO$_3$ substrate using XPEEM. By adjusting the bias voltage applied to the sample, a secondary photoelectron energy spectrum for each SAW phase (compressive and tensile) can be obtained, as done in Ref. \cite{foerster2019quantification}. The resulting difference between spectrums corresponds to the peak-to-peak piezoelectric voltage, $V_{\textrm{pp}}$. By solving the coupled elastic and electromagnetic equations, we can obtain the amplitude of the strain components, $e_{\textrm{xx}}$, $e_{\textrm{xz}}$ and $e_{\textrm{zz}}$, associated with the SAW images, which are listed in Table \ref{tab:strainRes}. It is important to note that the strain components may vary depending on SAW frequency. Thus, we measured the strain component in each sample at the same SAW excitation frequency and power that are used later on for magnetization imaging (see Table \ref{tab:strainRes})

\begin{table}[]
\caption{Amplitudes of the SAW strain components for each ample and SAW frequency}
\label{tab:strainRes}
{%
{\setlength{\extrarowheight}{3pt}%
\begin{tabular}{|c|ccc|l}
\cline{1-4}
Sample & $e_{xx} (10^{-5})$ & $e_{xy} (10^{-5})$ & $e_{zz} (10^{-5})$ &  \\ \cline{1-4}
Ni (1 GHz) & 14.3 & 2.95 & 4.23 &  \\
Ni (3 GHz) & 19.0 & 3.93 & 5.64 &  \\
Co (1 GHz) & 16.6 & 3.44 & 4.93 &  \\ \cline{1-4}
\end{tabular}}}
\end{table}

In the following sections, we show the results obtained from the XMCD imaging, and discuss the amplitude of the MAWs for the different samples and excitation frequencies.

\subsection{MAWs in Nickel}
The Ni sample was studied for two SAW frequencies: 1 and 3 GHz. Figures \ref{NiSWAmp}(a) and (b) show 2-phase XMCD images of excitations of both signals (1 and 3 GHz respectively). The bottom part of the images correspond to LiNbO$_3$ and displays the SAWs and the top part corresponds to Ni and shows the MAWs as indicated. Both images are taken with the same field-of-view ($20\times20$ $\mu$m) and the same color scale, to be able to distinguish the wavelength and the contrast difference. Fig.\ \ref{NiSWAmp}(c) shows the MAW profile of both frequencies with their corresponding sinusoidal fit to showcase the large amplitude difference (as well as the difference in wavelength). A summary of the normalized MAW amplitudes at different magnetic fields is presented in Fig. \ref{NiSWAmp}(d) for 1 GHz (black symbols, left axis) and 3 GHz (orange symbols, right axis). The overall magnetization precession amplitude at 1 GHz is one order of magnitude larger than that of 3 GHz while the measured strain amplitude at the substrate was similar for the two frequencies; this will be explored in more detail in Sec.\ \ref{section:discussion}.

\subsection{MAWs in Cobalt}
\begin{figure}[t]
    \centering
    \includegraphics[width=1\columnwidth]{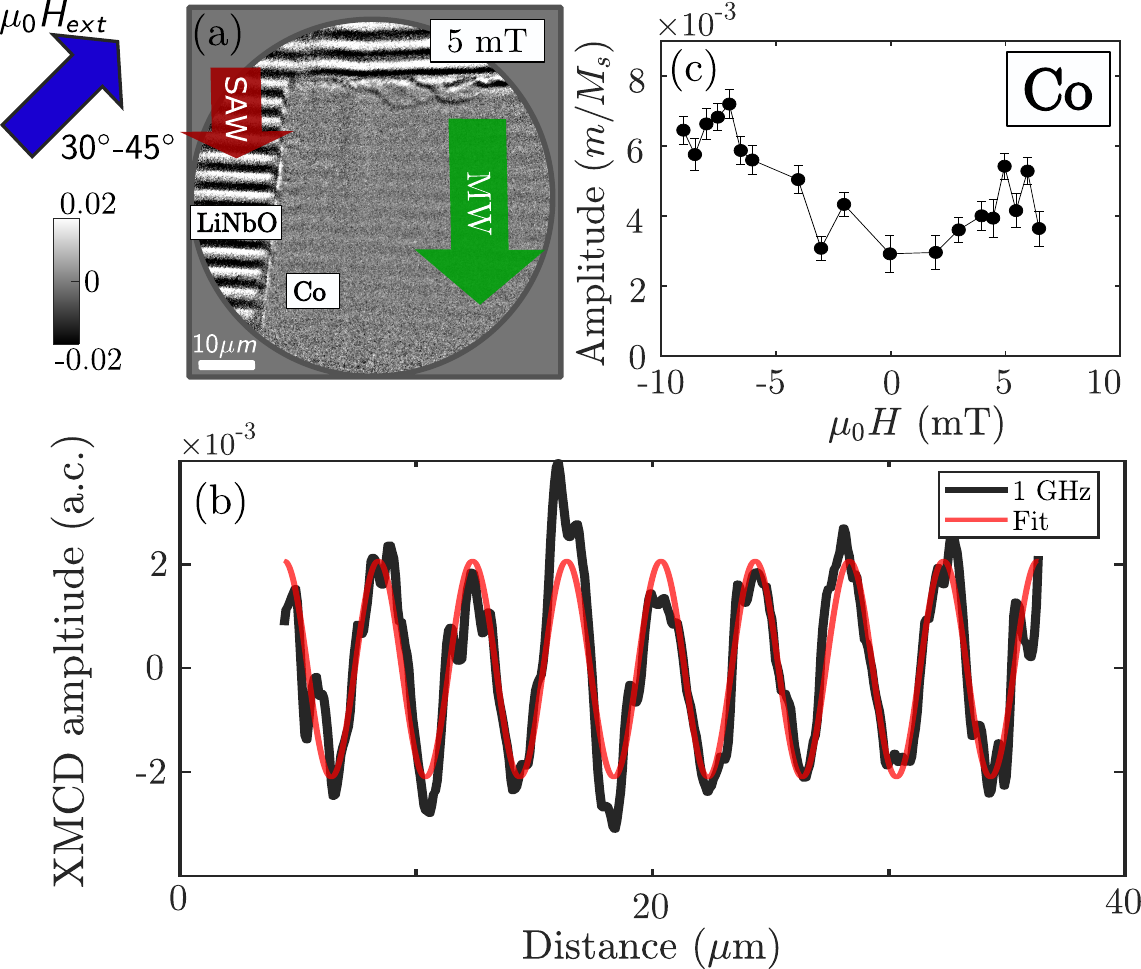}
    \caption{
    \textbf{(a)} XMCD image showing the piezoelectric, LiNbO$_3$, and the ferromagnetic sample, cobalt, as indicated at an applied magnetic field of 5 mT. The Co displays two big magnetic domains.
    \textbf{(b)} Experimentally determined MAW profile of the 5 mT XMCD image of a 1 GHz excitation (black curve) with the corresponding sinusoidal fit (red curve).
    \textbf{(c)} MAW amplitude as a function of the applied magnetic field excited by a 1 GHz SAW.
    }
    \label{CoSWAmp}
\end{figure}

The Co sample was only studied at 1 GHz. Figure \ref{CoSWAmp}(a) displays a representative 2-phase XMCD image with a field-of-view of $50\times50$ $\mu$m where clear SAWs in the LiNbO$_3$ and MAWs in the Co at 1 GHz are observed. The profile of this MAW is shown in Fig.\ \ref{CoSWAmp}(b) with the corresponding sinusoidal fit. Figure\ \ref{CoSWAmp}(c) collects results from different magnetic field values, with no clear trend and with an observable difference in amplitude for opposite direction of magnetic field, which could be the result of the magnetorotation coupling \cite{Alberto_2020, Xu_magnetorotation_2020,GiantNonReciprocalShah2020}. We notice here that the studied range of fields is small and one may hint a slight amplitude increase close to $\mu_0 H = \pm 7$ mT. The amplitude is two orders of magnitude smaller than Ni at 1 GHz (and even one order smaller than Ni at 3 GHz). We notice here that in both experiments the applied magnetic field formed an angle of $\sim 30-45^\circ$ with respect to the SAWs which could be outside the large attenuation region for Co (see Fig.\ \ref{NiCoFMR}b); we will discuss it in Sec.\ \ref{section:discussion}.

\section{Discussion}\label{section:discussion}
The experimental techniques employed here measured different quantities; a-FMR determines the attenuation of SAW energy under MAW resonance, while XMCD is a direct representation of the magnetization dynamics in the ferromagnetic film. By comparing our results with micromagnetic simulations we can determine the efficiency of converting strain into a magnetization variation, thus, obtaining the magnetoelastic constants. 

The experimental results demonstrate a notable consistency in the behavior of Ni and Co at 1 GHz in both a-FMR and XMCD. Starting with Ni, when placed under a magnetic field at $\varphi_H =$ 30-45$^\circ$ relative to the SAWs, the effect of the ME coupling is the strongest at approximately $\pm$4 mT and it decays as the magnetic field strength increases. However, while in a-FMR minimal attenuation is found beyond a magnetic field of 10 mT, XMCD still shows the presence of MAWs up to 20 mT. In the case of Co, the magnetic field that we could apply in XMCD imaging did not exceed 7 mT, limiting our knowledge at higher fields. However, a peak close to 5 mT is observed in both a-FMR and XMCD experiments, with a decrease in amplitude when approaching 0 mT. Note that the amplitude of the MAWs of Co is one and two orders of magnitude lower than nickel at 3 and 1 GHz, respectively (also seen in acoustic attenuation in Fig.\ \ref{NiCoFMR}). This phenomenon could be due to the high magnetization saturation of Co, which is three times larger than that of nickel. At 3 GHz frequency, XMCD data of Ni displays a similar trend with applied magnetic field as for the 1 GHz data, with a maximum around 10 mT.

As already discussed in Section \ref{section:results}, we determined the strain amplitude for each frequency and sample using XPEEM images of SAW on LiNbO$_3$ and employing the same power settings as in the MAW imaging, see  Table \ref{tab:strainRes}. We observed similar relative values between the strain components for both samples at 1 GHz since identical IDTs were used on the same LiNbO$_3$ substrates. Notably, we also found that the strain at the sample surface measured at 3 GHz in the Ni sample were comparable and, even, slightly larger than those at 1 GHz. Thus, the observed differences in MAW amplitudes cannot be explained with strain values. Therefore, we performed micromagnetic simulations to evaluate the overall internal magnetic fields and determine the expected MAW amplitudes at each frequency for both materials. We used MuMax3 \cite{mumax3}, which takes into account the free energy of different magnetic contributions, e.g. Zeeman, dipolar, exchange, and magnetoelastic to calculate an effective magnetic field. The contributions of each effect depend on the parameters of the experiment (SAW frequency and wavelength, external magnetic field, magnetization saturation). In particular, for the ME effect, the following free-energy expression is used.

\begin{align}\label{eq:Fme}
   \nonumber F_{\text{me}}= &\,\, B_1\left[\varepsilon_{x x} m_x^2+\varepsilon_{y y} m_y^2+\varepsilon_{z z} m_z^2\right] \\
    &+2 B_2\left[\varepsilon_{x y} m_x m_y+\varepsilon_{x z} m_x m_z+\varepsilon_{y z} m_y m_z\right],
\end{align}

\noindent where $B_1$ and $B_2$ are the magnetoelastic constants and are taken as parameters in our simulations, $m_i$ are the normalized magnetization components and $\varepsilon_{ij}$ are the strain components simulated as plane waves with the following expression:

\begin{equation}\label{eq:strainWaves}
    \varepsilon_{ij} = e_{ij}\exp(kx-\omega t),
\end{equation}

\noindent where $e_{ij}$ are the strain amplitudes of Table \ref{tab:strainRes}, $k$ is the wave vector of the SAW traveling in the x direction, and $\omega=2\pi f$ is the angular frequency of the SAWs. Since the SAW excited in the LiNbO3 substrate is a Rayleigh mode, only the strain amplitudes $e_{xx}$, $e_{zz}$, and $e_{xz}$ are non-zero in our experiment. 

\begin{table}[]

\label{tab:simProperties}
{\setlength{\extrarowheight}{3pt}%
\begin{tabular}{|l|ccc|l}
\cline{1-4}
Sample & $M_s$ (kA/m) & $A_{ex} (J/m)$ & $\alpha$ &  \\ \cline{1-4}
Ni & 490 &  $10^{-11}$  & 0.03 &  \\
Co & 1510 &  $10^{-11}$  & 0.007 &  \\ \cline{1-4}
\end{tabular}}
\caption{Sample properties used in the micromagnetic simulations. $M_s$ is the saturation magnetization, $A_{ex}$ the exchange and $\alpha$ the Gilbert damping.}
\end{table}

\begin{figure}[t]
    \centering
    \includegraphics[width=1\columnwidth]{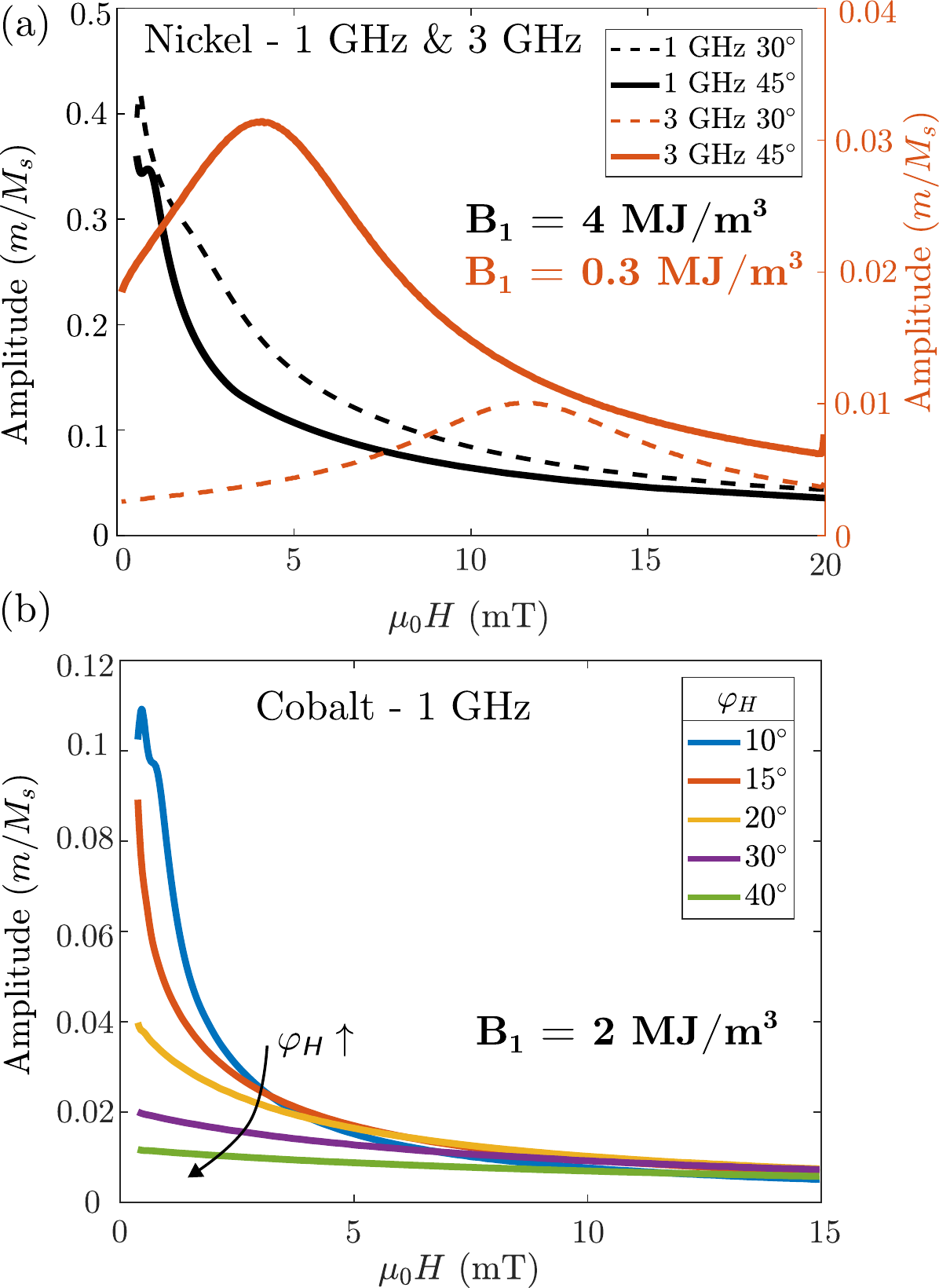}
    \caption{\textbf{(a)} Amplitude of MAW in Ni for excitation frequencies of 1 and 3 GHz with $\varphi_H=45^\circ$ (solid black and orange curves, respectively). The dashed lines provide the same information for a $\varphi_H=30^\circ$. The magnetoelastic constants used are written in the same color as the simulation curves. \textbf{(b)} Amplitude of the MAW in Co for an excitation frequency of 1 GHz. Each color represents a different magnetic field angle, with the arrow indicating the increase in the angle.}
    \label{NiCoSimulations}
\end{figure}

The simulations consist of sweeping the magnetic field gradually from 20 to 0 mT, mirroring the experimental procedure, and recording the magnetization values in time and space in presence of SAW. The choice of larger to lower magnetic fields is made to avoid magnetic switching and the formation of magnetic domains. The results in Fig.\ \ref{NiCoSimulations} show the amplitude of the MAWs determined by calculating the mean oscillating amplitude over time for each magnetic field. The results for Ni are presented in Fig.\ \ref{NiCoSimulations} (a) with MAW-amplitude for 1 GHz (black curves) represented on the left $y$-axis, and for 3 GHz (orange curves) represented on the right $y$-axis. Both frequencies were simulated at two different angles between the SAWs and the magnetic field ($\varphi_H$), at 45$^\circ$ (continuous lines) and 30$^\circ$ (dashed lines), to account for possible misalignment errors, and different magnetoelastic constants to match the experimental results. A first observation is that the maximum amplitude of MAWs shifts towards larger fields with increasing frequency---which has to do with ferromagnetic resonance fields that increase with excitation frequency. Another noticeable effect is the shift to lower fields of MAW maximum amplitude with increasing $\varphi_H$, which has previously been observed in a-FMR experiments \cite{weiler2011elastically,gowthamTravelingSurfaceSpinwave2015b}, and can be clearly seen in Fig.\ \ref{NiCoSimulations} (a) for the 3 GHz curves.  Figure \ref{NiCoSimulations} (b) illustrates the MAW amplitude of the Co sample when subjected to a 1 GHz excitation for different angles, $\varphi_H$, ranging from 10$^\circ$ to 40$^\circ$ (from blue to green curves). The misalignment between MAW and applied field has a more drastic impact on Co reducing the overall MAW amplitude. This occurs because the resonant frequency is shifted above the SAW frequency as $\varphi_H > 10^\circ$, resulting in no resonance.


We now focus on the experimental results showing a drop in efficiency at 3 GHz for Ni (see, Fig.\ \ref{NiSWAmp}). To account for the lower MAW amplitude, we varied the magnetoelastic constant, $B_1$, until the MAW amplitude matched the experimental values. For a 1 GHz excitation, a magnetoelastic constant of $B_1 = 4$ MJ/m$^3$ matched the experimental amplitude, while for a 3 GHz excitation, a reduced magnetoelastic constant of $B_1 = 0.3$ MJ/m$^3$ was needed. This loss in efficiency appeared unexpected to us, given that experiments are done in the same Ni layer and that physical processes involved in magneto-elasticity are expected to respond easily at the nanosecond scale. We notice here that the measured strain amplitudes were obtained at the piezoelectric surface and not at the ferromagnetic film. It could be possible that the reduced magnetic response at 3 GHz is related to a lower transmission of the SAW strain to the polycristalline thin layer under large SAW frequencies and short wavelengths. We measured AFM in our Ni films and the observed grains at the top of the film were not larger than a few nanometers, a value much shorter than the SAW wavelengths used in our experiment. However,  the interface between LiNbO$_3$ and Ni has not been characterized.


\begin{table}[]
\caption{Magnetoelastic constant values for each sample.}
\label{tab:MEct}
{%
{\setlength{\extrarowheight}{3pt}%
\begin{tabular}{|c|cc|}
\cline{1-3}
Sample & $B_1$ (MJ/m$^3$) at $\varphi_H$=45$^\circ$ &   \\ \cline{1-3}
Ni (1 GHz) & 4    &    \\
Ni (3 GHz) & 0.3  &    \\
Co (1 GHz) & 2    &    \\ \cline{1-3}
\end{tabular}}}
\end{table}

Overall, the results of our simulations agree quantitatively well with XMCD and a-FMR experiments for both Ni and Co. At 1 GHz SAW-excitation we obtained a value of $B_1$ for Ni of $\sim 4 $ MJ/m$^3$, comparable to previous experimental studies \cite{NiMagnetostrictionSong1994,IntroMagMatWiley2008, foerster2017direct,liSpinWaveGeneration2017b,casalsGenerationImagingMagnetoacoustic2020}. In Co, we obtained a magnetoelastic constant of $\sim 2$ MJ/m$^3$, which is 2 times smaller than the Ni one at 1 GHz. However, the simulations indicate that Co at smaller angles may achieve amplitudes approximately equal to 0.1, comparable to Ni, which could make Co an interesting material for straintronics, due to its low damping and high magnetization saturation. At 3 GHz SAW excitation, we found a reduction in the efficiency for the MAWs in Ni that requires decreasing the magnetoelastic constant (to $\sim 0.3$ MJ/m$^3$) approximately a factor of 10 compared with 1 GHz excitation. Although our model does not account for direct transfer of phonon angular momentum to magnetization \cite{Puebla_2020,Sasaki2021_phonon_spin}, which might be dependent on frequency, we consider that a plausible explanation for the drop in efficiency at 3 GHz could be related to a frequency dependence of the strain transfer efficiency between LiNbO3 and the Ni film.


\section{Conclusions}\label{section:conclusions}
We have studied the magnetoelastic effect in Ni and Co at 1 and 3 GHz using hybrid piezoelectric/ferromagnetic devices and SAWs. Our direct XPEEM-XMCD imaging shows that large-amplitude MAWs in Ni and Co at frequencies of 1 and 3 GHz can be directly imaged and quantified. We found a significant difference in MAW amplitude between Ni and Co at 1 GHz excitations; a difference that cannot be explained only by the larger magnetization saturation value of Co compared with the value of Ni and thus a lower value of magnetoelastic constant is obtained for Co. The found values for the magnetoelastic constants at 1 GHz are similar to those reported in the literature  \cite{foerster2017direct,liSpinWaveGeneration2017b,casalsGenerationImagingMagnetoacoustic2020,IntroMagMatWiley2008,NiMagnetostrictionSong1994}. Additional measurements at 3 GHz SAW excitations show a clear decay in MAW efficiency suggesting a drop in the phonon coupling between SAW at the piezoelectric substrate and SAW at the ferromagnetic thin film. Our study provides valuable insights into the coupling between strain and magnetization dynamics at the GHz frequency and may guide the development of more efficient acoustic spintronic (straintronic) devices.

\begin{acknowledgments}
The authors would like to thank N. Volkmer for their technical support in the preparation of the samples. MWK acknowledges Marie Skłodowska-Curie grant agreement No. 754397 (DOC-FAM) from EU Horizon 2020. MR, BC, JMH, and FM acknowledge funding from MCIN/AEI/10.13039/501100011033 through grant number: PID2020-113024GB-100. MWK, MF, and MAN acknowledge funding from MCIN through grant number: PID2021-122980OB-C5. LA, MF and SRG acknowledge funding from MCIN through grant number: RTI2018-095303. SRG also acknowledges funding from Marie Sklodowska-Curie grant number 101061612.
\end{acknowledgments}
ç

\begin{thebibliography}{10}

\bibitem{locatelliSpintorqueBuildingBlocks2014}
N.~Locatelli, V.~Cros, and J.~Grollier,
\newblock Nature Materials {\bf 13}, 11 (2014).

\bibitem{RALPH20081190}
D.~Ralph and M.~Stiles,
\newblock Journal of Magnetism and Magnetic Materials {\bf 320}, 1190 (2008).

\bibitem{brataasCurrentinducedTorquesMagnetic2012}
A.~Brataas, A.~D. Kent, and H.~Ohno,
\newblock Nature Materials {\bf 11}, 372 (2012).

\bibitem{Bukharaev2018straintornicRev}
A.~A. Bukharaev, A.~K. Zvezdin, A.~P. Pyatakov, and Y.~K. Fetisov,
\newblock Physics-Uspekhi {\bf 61}, 1175 (2018).

\bibitem{Foerster_2019}
M.~Foerster and F.~Macià,
\newblock J. Phys. Condens. Matter {\bf 31}, 190301 (2019).

\bibitem{Mangin2014}
C.-H. Lambert et~al.,
\newblock Science {\bf 345}, 1337 (2014).

\bibitem{Barangi2015StraintronicsPowerEff}
M.~Barangi and P.~Mazumder,
\newblock IEEE Nanotechnology Magazine {\bf 9}, 15 (2015).

\bibitem{wang2014magrev_strain}
Z.~Wang et~al.,
\newblock ACS Nano {\bf 8}, 7793 (2014),
\newblock PMID: 25093903.

\bibitem{Huang2014magrev_strain}
H.~B. Huang, J.~M. Hu, T.~N. Yang, X.~Q. Ma, and L.~Q. Chen,
\newblock Applied Physics Letters {\bf 105}, 122407 (2014).

\bibitem{cenkerReversibleStraininducedMagnetic2022}
J.~Cenker et~al.,
\newblock Nature Nanotechnology {\bf 17}, 256 (2022).

\bibitem{gangulyMagnetoelasticSurfaceWaves1976}
A.~K. Ganguly, K.~L. Davis, D.~C. Webb, and C.~Vittoria,
\newblock Journal of Applied Physics {\bf 47}, 2696 (1976).

\bibitem{fengMechanismInteractionSurface1982}
I.-a. Feng, M.~Tachiki, C.~Krischer, and M.~Levy,
\newblock Journal of Applied Physics {\bf 53}, 177 (1982).

\bibitem{yang2021acoustic}
W.-G. Yang and H.~Schmidt,
\newblock Applied Physics Reviews {\bf 8}, 021304 (2021).

\bibitem{puebla2022perspectives}
J.~Puebla, Y.~Hwang, S.~Maekawa, and Y.~Otani,
\newblock Applied Physics Letters {\bf 120}, 220502 (2022).

\bibitem{casalsGenerationImagingMagnetoacoustic2020}
B.~Casals et~al.,
\newblock Physics Review Letters {\bf 124}, 137202 (2020).

\bibitem{weiler2011elastically}
M.~Weiler et~al.,
\newblock Physical Review Letters {\bf 106}, 117601 (2011).

\bibitem{gowthamTravelingSurfaceSpinwave2015b}
P.~G. Gowtham, T.~Moriyama, D.~C. Ralph, and R.~A. Buhrman,
\newblock Journal of Applied Physics {\bf 118}, 233910 (2015).

\bibitem{Labanowski2016}
D.~Labanowski, A.~Jung, and S.~Salahuddin,
\newblock Applied Physics Letters {\bf 108}, 022905 (2016).

\bibitem{seemann2022magnetoelastic}
K.~M. Seemann et~al.,
\newblock Physical Review B {\bf 105}, 144432 (2022).

\bibitem{Kuszewski_2018}
P.~Kuszewski et~al.,
\newblock Journal of Physics: Condensed Matter {\bf 30}, 244003 (2018).

\bibitem{McCord_AEM_2022}
C.~Müller et~al.,
\newblock Advanced Electronic Materials {\bf 8}, 2200033 (2022).

\bibitem{rovirola2023_physrevapp}
M.~Rovirola et~al.,
\newblock Physics Review Applied {\bf 20}, 034052 (2023).

\bibitem{khaliq2023antiferromagnetic}
M.~W. Khaliq et~al.,
\newblock Magneto-acoustic waves in antiferromagnetic cumnas excited by surface
  acoustic waves, 2023.

\bibitem{SAW_roadmap_2019}
P.~Delsing et~al.,
\newblock Journal of Physics D: Applied Physics {\bf 52}, 353001 (2019).

\bibitem{Zhang2020MOKEspinwaves}
D.-L. Zhang et~al.,
\newblock Science Advances {\bf 6}, eabb4607 (2020).

\bibitem{qinNanoscaleMagnonicFabryPerot2021}
H.~Qin et~al.,
\newblock Nature Communications {\bf 12}, 2293 (2021).

\bibitem{foerster2017direct}
M.~Foerster et~al.,
\newblock Nature Communications {\bf 8}, 407 (2017).

\bibitem{foersterSubnanosecondMagnetizationDynamics2018}
M.~Foerster, L.~Aballe, J.~M. Hern{\`a}ndez, and F.~Maci{\`a},
\newblock MRS Bulletin {\bf 43}, 854 (2018).

\bibitem{Slobodnik1976_SAW}
A.~Slobodnik,
\newblock Proceedings of the IEEE {\bf 64}, 581 (1976).

\bibitem{AballeALBA2015}
L.~Aballe, M.~Foerster, E.~Pellegrin, J.~Nicolas, and S.~Ferrer,
\newblock Journal of Synchrotron Radiation {\bf 22}, 745 (2015).

\bibitem{foerster2019quantification}
M.~Foerster et~al.,
\newblock Journal of Synchrotron Radiation {\bf 26}, 184 (2019).

\bibitem{Waqas2023}
M.~W. Khaliq et~al.,
\newblock Ultramicroscopy {\bf 250}, 113757 (2023).

\bibitem{Puebla_2020}
J.~Puebla et~al.,
\newblock Journal of Physics D: Applied Physics {\bf 53}, 264002 (2020).

\bibitem{dreherSurfaceAcousticWave2012}
L.~Dreher et~al.,
\newblock Physical Review B {\bf 86}, 134415 (2012).

\bibitem{spinPumpingCoWeiler2012}
M.~Weiler et~al.,
\newblock Physics Review Letters {\bf 108}, 176601 (2012).

\bibitem{Alberto_2020}
A.~Hern\'andez-M\'{\i}nguez, F.~Maci\`a, J.~M. Hern\`andez, J.~Herfort, and
  P.~V. Santos,
\newblock Physics Review Applied {\bf 13}, 044018 (2020).

\bibitem{Xu_magnetorotation_2020}
M.~Xu et~al.,
\newblock Science Advances {\bf 6}, eabb1724 (2020).

\bibitem{GiantNonReciprocalShah2020}
P.~J. Shah et~al.,
\newblock Science Advances {\bf 6}, eabc5648 (2020).

\bibitem{mumax3}
A.~Vansteenkiste et~al.,
\newblock AIP Advances {\bf 4}, 107133 (2014).

\bibitem{NiMagnetostrictionSong1994}
O.~Song, C.~A. Ballentine, and R.~C. O'Handley,
\newblock Applied Physics Letters {\bf 64}, 2593 (1994).

\bibitem{IntroMagMatWiley2008}
B.~Cullity and C.~Graham,
\newblock {\em Magnetostriction and the Effects of Stress},
\newblock John Wiley \& Sons, Ltd, 2008.

\bibitem{liSpinWaveGeneration2017b}
X.~Li, D.~Labanowski, S.~Salahuddin, and C.~S. Lynch,
\newblock J. Appl. Phys. {\bf 122}, 043904 (2017).

\bibitem{Sasaki2021_phonon_spin}
R.~Sasaki, Y.~Nii, and Y.~Onose,
\newblock Nature Communications {\bf 12}, 2599 (2021).

\end{thebibliography}

\end{document}